\documentclass[12pt]{article}   
\usepackage{amsmath}
\usepackage{amssymb}
\usepackage{graphicx}
\begin{document}
\title{Simple kinematic effects on de Sitter expanding universe}

\author{Ion I. Cot\u aescu %\inst{1}
\thanks{e-mail: i.cotaescu@e-uvt.ro}\\
{West University of Timi\c soara,} \\{V. P\^ arvan Ave. 4, RO-300223, Timi\c soara, Romania}}

\maketitle
\begin{abstract}
Our recent results concerning the transformation under isometries of the conserved quantities on de Sitter manifolds, allow us to define the rest frame and study the relative geodesic motion in terms of conserved momentum, revealing thus various relativistic effects [I. I. Cotaescu, arXiv:1701.08499]. Here we discuss  the time dilation (of the twin paradox) and the Lorentz contraction on the de Sitter expanding portions, pointing out the dependence of these effects on the position and moment in which the time and lengths are measured. 
\end{abstract} 

PACS: {04.20.Cv} and { 04.62.} % end of PACS codes

\section{Introduction}
\label{intro}

In special and general relativity there are three maximally symmetric spacetimes, the Minkowski, de Sitter and Anti-de Sitter ones \cite{SW}, whose isometry groups have the same number of parameters but different structures such that only on the Minkowski and de Sitter spacetime we meet translations and, consequently, conserved momenta.    

In Minkowski spacetime the translations play a crucial role in Wigner's theory of the induced representations of the Poincar\' e group \cite{Wig} since the classification of the unitary irreducible representations \cite{Mc} is based on the orbital analysis in the energy-momentum space. Moreover, the equivalence among the covariant representations and the unitary ones may be studied exclusively in momentum representation \cite{WKT}. 

Unfortunately, this method cannot be applied  to the de Sitter isometries since here the momentum is combined with other conserved quantities that depend on coordinates, transforming among themselves under isometries \cite{C1}. Therefore, in this case we cannot speak about the energy-momentum space and  its orbits.  Nevertheless, despite of this difficulty, we may study how different observers measure the conserved quantities on geodesics resorting to our previous methods in investigating external symmetries and the corresponding conserved quantities \cite{C1,ES,C2,C3,C4}. 

In Ref. \cite{Cnew} we solved completely the problem  of the relative motion on de Sitter spacetimes considering the (conformal) Euclidean and de Sitter-Pailev\' e local charts as inertial natural frames where each geodesic is determined by the conserved momentum  in a certain position at a given moment.  Obviously, the coordinates of these local charts as well as the conserved quantities on geodesics are related among themselves  through de Sitter isometries that take over the role of the Poincar\' e isometries in special relativity. The main point of our approach is the parametrization of these isometries in terms of conserved momenta and the definition of the rest frames,  obtaining thus a framework in which the relative motion can be studied as in the flat case. 

Among the examples briefly outlined in Ref. \cite{Cnew} the time dilation and Lorentz contraction on the de Sitter expanding portion were studied only in the particular case when the time and length are measured near the origin of the mobile frame. Here we would like to continue this study, analyzing how these effects depend on the relative position of the measurement point with respect to the mobile frame.  After we briefly present in the next two sections the de Sitter geodesics and the isometries relating the fixed and mobile frames, we study the mentioned relativistic effects in the last section.

\section{de Sitter geodesics}

The de Sitter spacetime $(M,g)$ is defined as the hyperboloid of radius $1/\omega$  in the five-dimensional flat spacetime $(M^5,\eta^5)$ of coordinates $z^A$  (labeled by the indices $A,\,B,...= 0,1,2,3,4$) having the Minkowskian metric $\eta^5={\rm diag}(1,-1,-1,-1,-1)$. The local charts $\{x\}$  of coordinates $x^{\mu}$ ($\alpha,\mu,\nu,...=0,1,2,3$) can be introduced on $(M,g)$ giving the set of functions $z^A(x)$ which solve the hyperboloid equation,
\begin{equation}\label{hip}
\eta^5_{AB}z^A(x) z^B(x)=-\frac{1}{\omega^2}\,.
\end{equation}
where  $\omega$ denotes the Hubble de Sitter constant since in our notations  $H$ is reserved for the energy (or Hamiltonian) operator \cite{C1}. In what follows we consider the local charts with de Sitter-Painlev\' e coordinates $\{t, \vec{x}\}$ giving the line element 
\begin{equation}
ds^2=(1-\omega^2 {\vec{x}}^2)dt^2+2\omega \vec{x}\cdot d\vec{x}\,dt -d\vec{x}\cdot d\vec{x}\,, 
\end{equation}
on the expanding portion of the de Sitter spacetime.

These charts which play here the role of inertial frames can be transformed among themselves the de Sitter isometries given by the gauge group $G(\eta^5)=SO(1,4)$ of the embedding manifold $(M^5,\eta^5)$  that leave  invariant its metric and implicitly Eq. (\ref{hip}). Therefore, given a system of coordinates defined by the functions $z=z(x)$, each transformation ${\frak g}\in SO(1,4)$ defines the isometry $x\to x'=\phi_{\frak g}(x)$ derived from the system of equations $z[\phi_{\frak g}(x)]={\frak g}z(x)$. In this framework,  we may study the relativistic effects measured in different local charts if we know the isometries relating these charts. 

An observer staying at rest in the origin $O$ of the chart  $\{t, \vec{x}\}$ measuring different geodesics, draws the conclusion that any timelike geodesic $\vec{x}=\vec{x}(t)$ may be the trajectory of a freely falling particle of mass $m$ determined by the initial condition $\vec{x}(t_0)=\vec{x}_0$ and the conserved momentum $\vec{P}$  of this particle. The geodesic equation with these parameters reads
\begin{equation}\label{geod}
\vec{x}(t)=\vec{x}_0\,e^{\omega (t-t_0)}+\frac{\vec{n}_P e^{\omega t}}{\omega P}\left(\sqrt{m^2 +P^2e^{-2\omega t_0}}-\sqrt{m^2 +P^2e^{-2\omega t}}\right)
\end{equation}
where we denoted $P=|\vec{P}|$ and $\vec{n}_P=\frac{\vec{P}}{P}$. Another important conserved quantity along this geodesic is the energy depending on the initial condition and momentum as 
\begin{equation}
E=\omega \vec{P}\cdot \vec{x}_0\,e^{-\omega t_0}+\sqrt{m^2 +P^2e^{-2\omega t_0}}\,.
\end{equation}
Notice that the point $(t_0,\vec{x}_0)$ is arbitrary chosen on geodesic such that this can be replaced at any time with any current point  $(t,\vec{x}(t))$ without changing the value of $E$.

In the particular case when the geodesic across the origin $\vec{x}_0=0$ at time $t_0=0$, we obtain $E=\sqrt{m^2 +P^2}$ while the velocity $\vec{v}(t)=\dot{\vec{x}}(t)$ at $t=0$ is related to momentum and energy as  
\begin{equation}\label{v0}
\vec{v}_0=\vec{v}(0)=\frac{\vec{P}}{E} \to \vec{P}=\frac{m \vec{v}_0}{\sqrt{1-\vec{v}_0\,^2}}\,,\quad E=\frac{m }{\sqrt{1-\vec{v}_0\,^2}}\,,
\end{equation}
recovering thus familiar formulas from special relativity.

\section{Relative geodesic motion}

In Ref. \cite{Cnew} we have shown that we can define the rest frame $\{t', {\vec{x}\,}'\}$ of the mobile $m$ as the local chart in which this stays at rest in the origin $ {\vec{x}\,}'=0$. This means that for an observer $O'$, comoving with the rest frame, the mobile $m$ has the momentum $\vec{P}=0$. 

{ \begin{figure}
    \centering
    \includegraphics[scale=.50]{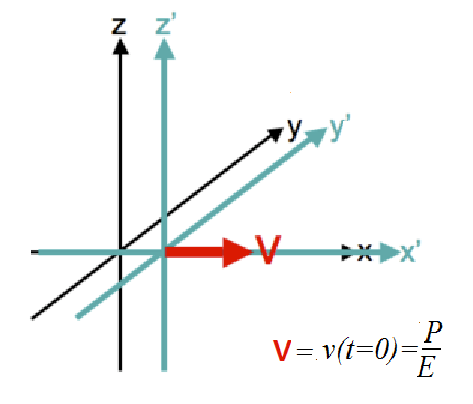}
    \caption{Initial conditions: the origins coincide at $t=t'=0$.}
  \end{figure}}
  
We demonstrated that if the clocks of the observers $O$ and $O'$ are synchronized according to the natural initial conditions (as in Fig. 1),
\begin{equation}\label{icon1}
t_0=t'_0=0\,, \quad \vec{x}(t_0)={\vec{x}\,}'(t_0')=0\,,
\end{equation}
then we can write simply the isometry $x=\phi_{{\frak g}(\vec{P})}(x')$ between these frames. Thus the transformations of this direct isometry read
\begin{eqnarray}
t(t',{\vec{x}}')&=&\frac{1}{\omega}\, \ln\left( e^{\omega t'}+\frac{\omega}{m}{\vec{x}}'\cdot \vec{P} + \frac{E-m}{m}\,\omega\Theta'  \right)\,,\label{Eq1A}\\
{\vec{x}}(t',{\vec{x}}')&=&{\vec{x}}'+\frac{\vec{P}}{m}\left(\frac{{\vec{x}}'\cdot\vec{P}}{E+m}+\Theta'\right)\,,\label{Eq2A}
\end{eqnarray}
where we used the identity $m^2+ P^2=E^2$ that holds in this case and denote 
\begin{equation}
\Theta'=\frac{1}{2\omega}\left[e^{\omega t'}-e^{-\omega t'}(1-\omega^2 {{\vec{x}\,}'}^2)\right]\,.
\end{equation}
These isometries may be used in applications since the de Sitter-Painlev\' e coordinates are suitable for physical interpretations. For this reason we write explicitly the inverse isometry $x'=\phi_{{\frak g}(-\vec{P})}(x)$ which has the transformation rules
\begin{eqnarray}
t'(t,{\vec{x}})&=&\frac{1}{\omega}\, \ln\left( e^{\omega t}-\frac{\omega}{m}{\vec{x}}\cdot \vec{P} + \frac{E-m}{m}\,\omega\Theta  \right)\,,\label{Eq1B}\\
{\vec{x}}'(t,{\vec{x}})&=&{\vec{x}}+\frac{\vec{P}}{m}\left(\frac{{\vec{x}}\cdot\vec{P}}{E+m}-\Theta\right)\,,\label{Eq2B}
\end{eqnarray}
where now we denote  
\begin{equation}
\Theta=\frac{1}{2\omega}\left[e^{\omega t}-e^{-\omega t}(1-\omega^2 {{\vec{x}}}^2)\right]\,.
\end{equation}

For small values of $\omega$ we may consider the expansions 
\begin{eqnarray}
t&=&\frac{E}{m}\,t' +\frac{{\vec{x}\,}'\cdot \vec{P}}{m}+\frac{1}{2m^2}\left[m(E-m)
{{\vec{x}\,}'}^2+m^2 {t'\,}^2\right. \nonumber\\
&&\hspace*{24mm}\left.-(Et'+\vec{P}\cdot{\vec{x}\,}')^2\right]\omega +{\cal O}(\omega^2)\,,\\
{\vec{x}}&=&{\vec{x}\,}'+\frac{\vec{P}}{m}\left[\frac{{\vec{x}\,}'\cdot\vec{P}}{E+m}+t' \right]
+\frac{1}{2m}{{\vec{x}\,}'}^2 \vec{P}\, \omega +{\cal O}(\omega^2)\,,
\end{eqnarray}
instead of Eqs. (\ref{Eq1A}) and (\ref{Eq2A}). We obtain thus the corrections of the first order and verify that for  $\omega=0$ we recover  just to the usual Lorentz transformations between the rest frame of a mobile $m$ of momentum $\vec{P}$ and the frame of the fixed observer  in  Minkowski spacetime.

\section{Relativistic effects}

The isometries studied here open the door to a large field of applications from the elementary relativistic effects up to the study of the properties of the covariant fields. 

In what follows  we focus on the relative motion of $m$ in the local chart $O$ with de Sitter-Painlev\' e coordinates for which we use the initial conditions  (\ref{icon1}). In order to avoid confusion we denote the space coordinates of $m$ in this frame by  $\vec{x}_*(t)$ and write the geodesic equation 
\begin{equation}\label{geodP}
\vec{x}_*(t)=\frac{\vec{P}}{\omega {P}^2}\,\left(Ee^{\omega t}-\sqrt{ m^2 e^{2\omega t}+{P}^2}\, \right)\,,
\end{equation}
resulted straightforwardly from Eqs. (\ref{geod}). The covariant four-velocity  can be derived  as
\begin{eqnarray}
u^0_*&=&\frac{dt}{ds}=\sqrt{1+\frac{P^2}{m^2}\,e^{-2\omega t}}\,,\\
u^i_*&=&\frac{dx^i_*}{ds}=\frac{P_i}{m}e^{-\omega t}+\omega x^i_* (t)\sqrt{1+\frac{P^2}{m^2}\,e^{-2\omega t}}\,,
\end{eqnarray}
laying out the relation between the covariant momentum $p^{\mu}=mu^{\mu}$ and the conserved one. Hereby we observe that for large values of $t$ we recover the well-known Hubble law since then $u^0 \to 1$ and $u^i \to \omega x^i(t)$ where   $\vec{x}(t) \to \vec{P}[\omega(E+m)]^{-1} e^{\omega t}$\,.
  
\begin{figure}
    \centering
    \includegraphics[scale=.41]{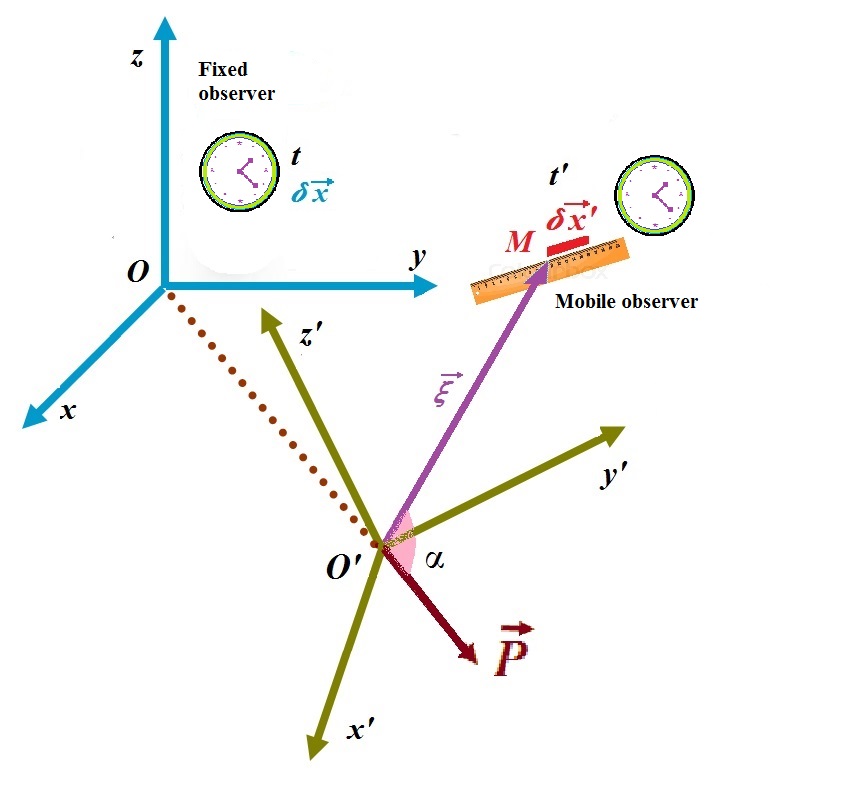}
    \caption{Measuring the time dilation and Lorentz contraction }
  \end{figure}
  
The applications we intend to study here are the simple kinematic effects of time dilation (observed in the so called twin paradox) and the Lorentz contraction which are strongly dependent on the position where the time and length are measured. Therefore, we assume that the measurement is performed in the point $M$ of coordinates ${\vec{x}}'=\vec{\xi}$, fixed rigidly to the mobile frame $O'$.  

Under such circumstances,  we may exploit the general relations 
\begin{eqnarray}
\delta t &=&\left.\frac{\partial t(t',{\vec{x}}')}{\partial t'}\right| _{{\vec{x}}'=0}\delta t' + \left.\frac{\partial t(t',{\vec{x}}')}{\partial x^{\prime\, i}}\right| _{{\vec{x}}'=0}\delta  x^{\prime\, i}\,,\label{cucu}\\
\delta x^j &=&\left.\frac{\partial x^j(t',{\vec{x}}')}{\partial t'}\right| _{{\vec{x}}'=0}\delta t' +\left.\frac{\partial x^j(t',{\vec{x}}')}{\partial x^{\prime\, i}}\right| _{{\vec{x}}'=0}\delta  x^{\prime\, i}\,,\label{mucu}
\end{eqnarray} 
among the quantities $\delta t, \delta x^j$ and $\delta t',\delta x^{\prime\, j}$ supposed to be measured by the observers $O$ and respectively $O'$.  

First we consider a clock in $M$ indicating $\delta t'$ without changing its position such that $\delta x^{\prime\, i}=0$. Then,  after a little calculation, we obtain the time dilation observed by $O$,    
\begin{equation}
\delta t (t)=\delta t' \beta(t)
\end{equation}
where we denote 
\begin{equation}
\beta(t)=\sqrt{[1-\omega\, \vec{\xi}\cdot \vec{\kappa}(t)]^2+\vec{\kappa}(t)^2\left(1-\omega^2 {\vec{\xi}\,}^2\right)}\,, \quad \vec{\kappa}(t)=\frac{\vec{P}}{m}e^{-\omega t}\,.
\end{equation}
Similarly but with the supplemental simultaneity condition $\delta t=0$ in Eq. (\ref{cucu}) and substituting  $\delta t'$ in Eq. (\ref{mucu}) we derive the general formula of the Lorentz contraction  that reads
\begin{equation}
\delta \vec{x}(t)=\delta{\vec{x}\,}'+\vec{\kappa}(t)\left\{ \frac{\delta{\vec{x}\,}'}{\beta(t)}\cdot\left[\omega\, \vec{\xi}-\frac{\left(1-\omega^2 {\vec{\xi}}^2\right) \vec{\kappa}(t)}{1+\beta(t)-\omega\, \vec{\xi}\cdot \vec{\kappa}(t)}\right]\right\}\,.
\end{equation}
These general results are encapsulated in complicated formulas such that it deserves to discuss some consequences.

We can verify first that the function $\beta$ is well-defined for all the values of ${\xi}\le \frac{1}{\omega}$. The problem is to establish the conditions in which this function satisfy $\beta>1$ giving a real time dilation.  In what concerns the Lorentz contraction we observe that this is strongly dependent from the relative positions of the vectors $\delta{\vec{x}}\,'$, $\vec{\xi}$ and $\vec{P}$. Therefore, it is convenient to consider explicitly the angle $\alpha$ between $\vec{\xi}$ and $\vec{P}$ and to split  $\delta{\vec{x}}\,'$ in the component $\delta{x}'_{||}$ parallel to $\vec{n}_P$ and the orthogonal one, $\delta{x}'_{\perp}$. Then the previous results can be rewritten as 
\begin{eqnarray}
\beta(t)&=&\sqrt{\left[1-\omega\, {\xi}\,{\kappa}(t)\cos \alpha\right]^2+{\kappa}(t)^2\left(1-\omega^2 {{\xi}\,}^2\right)}\,,\label{B1}\\
\delta x_{||}(t)&=&\delta{x'}_{||}
+\frac{\kappa(t)}{\beta(t)}\omega\xi \left( \delta{x}'_{\perp}\sin \alpha+ \delta {x'}_{||} \cos\alpha \right)\nonumber\\
&&~~~~~~-\frac{\kappa(t)^2}{\beta(t)}\frac{\left(1-\omega^2 {\xi}^2\right) }{1+\beta(t)-\omega\, {\xi}{\kappa}(t)\cos\alpha}\delta {x'}_{||}\,,\label{B2}\\
\delta{x}_{\perp}(t)&=&\delta{{x}\,}_{\perp}'\label{B3}
\end{eqnarray} 
where now ${\kappa}(t)=\frac{{P}}{m}\,e^{-\omega t}$

Hereby we see that the principal feature of these effects on de Sitter manifold is  that  these are decreasing in time vanishing  in the limit of $t \to \infty$ when $\vec{\kappa}(t)\to 0$ and $\beta  (t) \to 1$.

 \begin{figure}
    \centering
    \includegraphics[scale=.40]{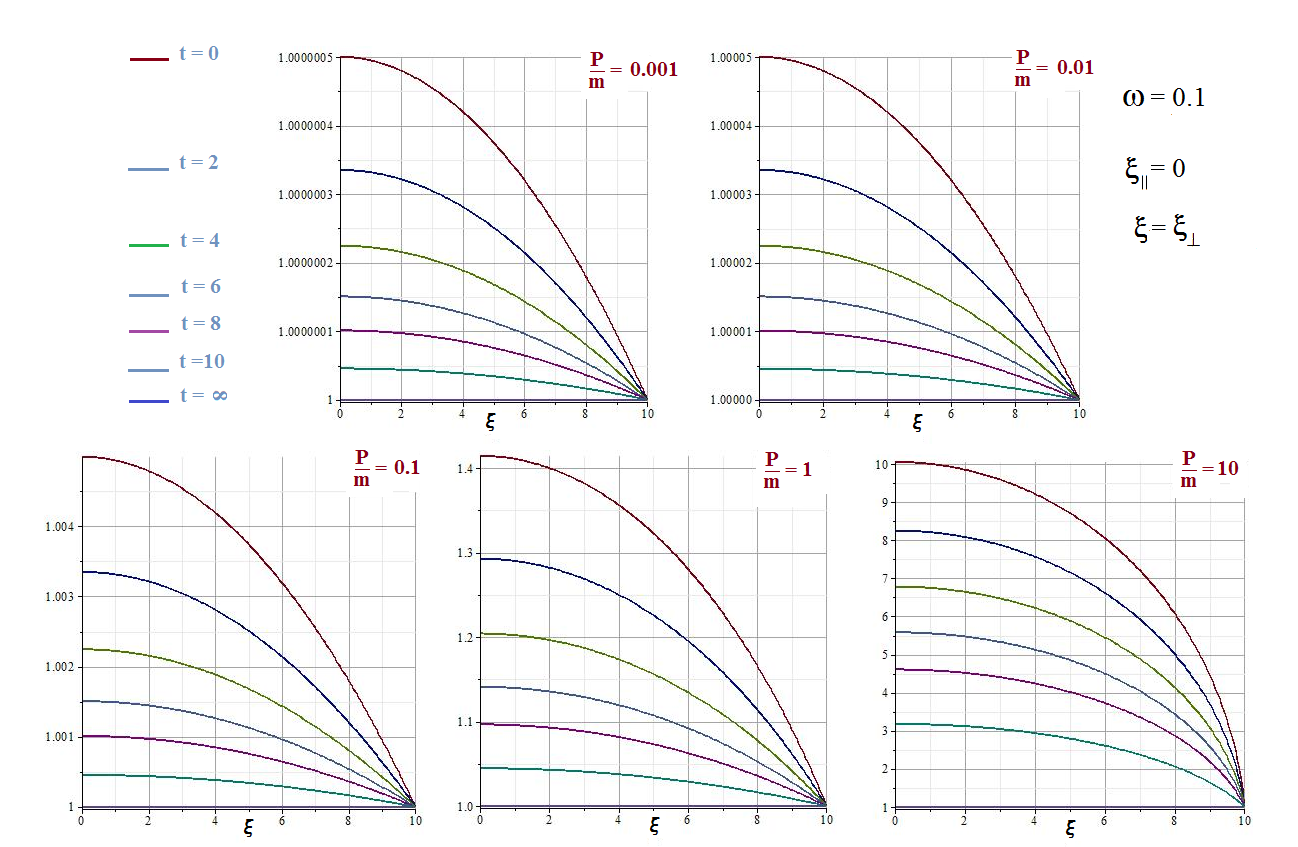}
    \caption{Function $\beta$ versus $\xi=\xi_{\perp}= |\vec{\xi}|$ when $\xi_{||}=\vec{\xi}\cdot\vec{n}_P=0$. }
  \end{figure}

\begin{figure}
    \centering
    \includegraphics[scale=.40]{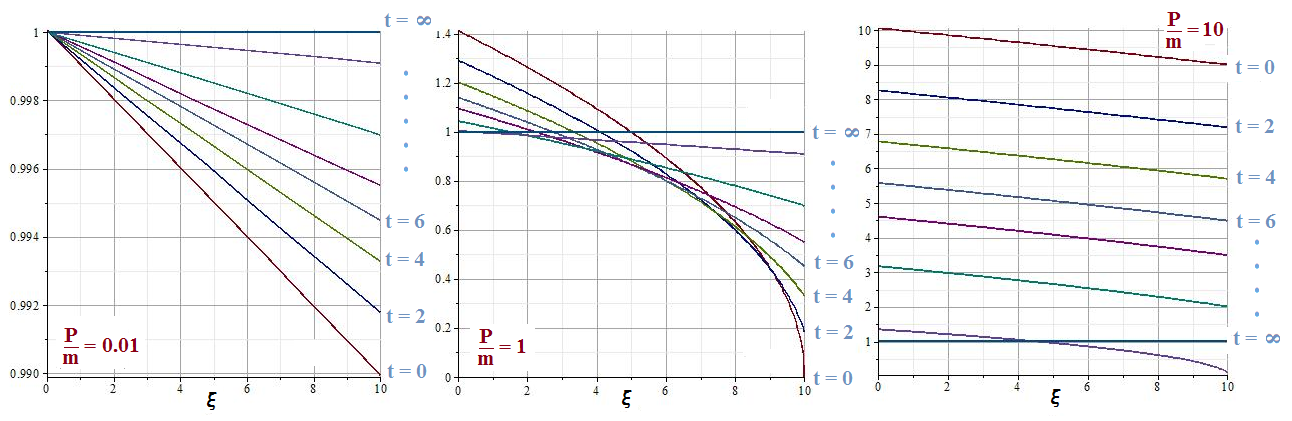}
    \caption{Function $\beta$ versus $\xi=\xi_{||}$ when $\xi_{\perp}=0$. }
  \end{figure}
  
In general, the effects governed by Eqs. (\ref{B1}), (\ref{B2}) and (\ref{B3}) are complicated with different behaviors in two extreme cases, namely when $\vec{\xi}$ is (i) orthogonal to $\vec{n}_P$ or (ii)  parallel to it. In the first case the function $\beta>1$ assures the expected time dilation but which is decreasing when $\xi$ increases (see Fig. 3). The case (ii) is more interesting because of the function $\beta$ which can take even values less than one, transforming the expected dilation in a time contraction. This happens for small values of the momentum of the mobile frame as shown  in Fig. 4.  

The Lorentz contraction is complicated and deserves a special study. Here we restrict ourselves to illustrate how the image depends on the position of the objects in the mobile frame  showing how the fixed observer $O$ see the motion of a pair of spheres moving with the mobile frame without changing their relative position (Fig. 5).

A particular case studied in Ref. \cite{Cnew} is when the measurement is performed in the origin of the mobile frame, $M\equiv O'$ (with $\vec{\xi}=0$ and $\vec{\kappa}=\vec{\kappa}(t_*)$). Then the time dilation takes the simple form
\begin{equation}
\delta t=\delta t' \left(1+\frac{P^2}{m^2}e^{-2\omega t}\right)^{\frac{1}{2}}\,,
\end{equation}   
while the Lorentz contraction along the direction of unit vector $\vec{n}_P$ becomes
\begin{equation}
\delta x_{||}=\delta x'_{||}\left(1+\frac{P^2}{m^2}e^{-2\omega t}\right)^{-\frac{1}{2}}\,.
\end{equation}  
It is remarkable that hereby we recover  the property $\delta t \delta x_{||}=\delta t' \delta x'_{||}$ known from the special relativity.   

 \begin{figure}
    \centering
    \includegraphics[scale=.40]{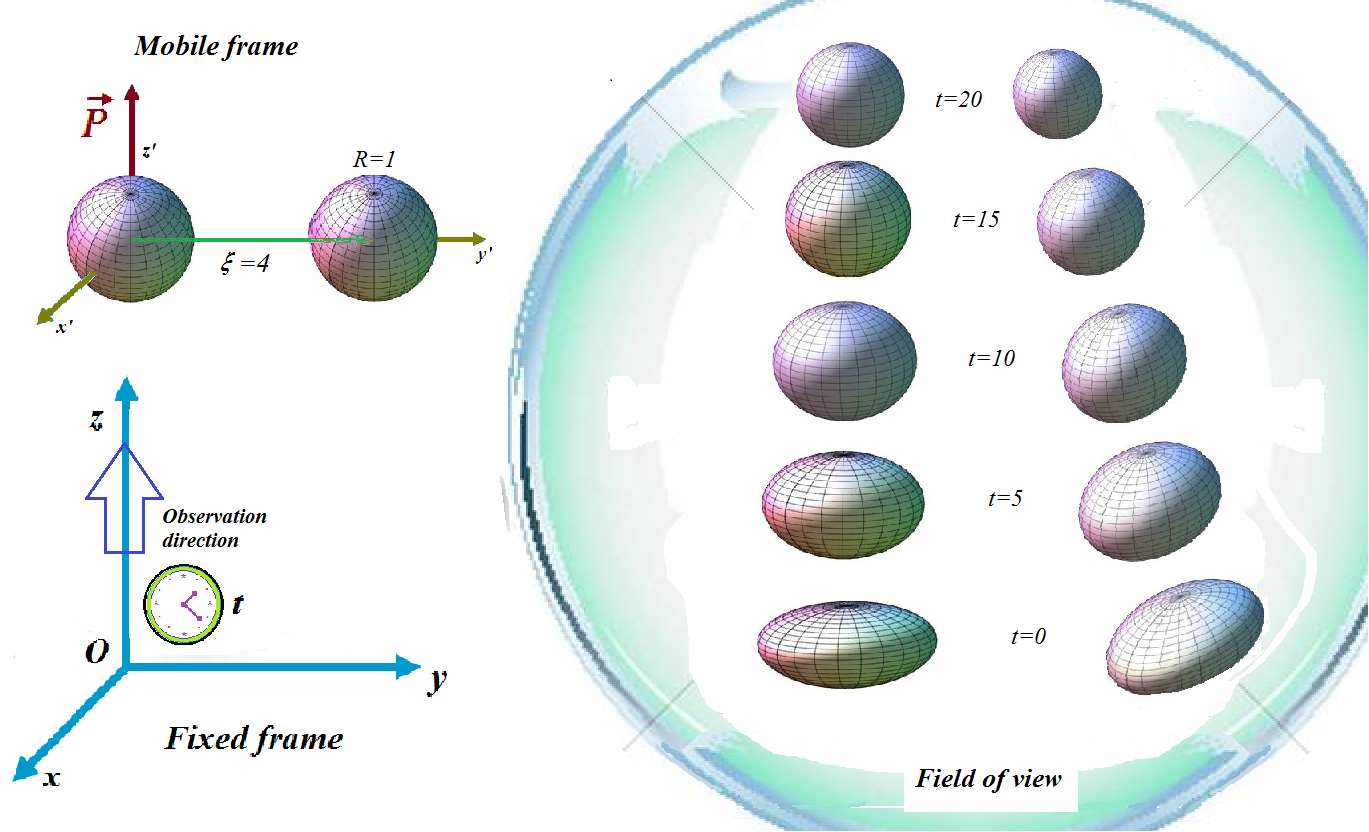}
    \caption{Flying spheres moving and observed along the $z$ axis. The successive images lay out the Lorentz contraction observed for the central sphere and the deformation of the lateral one. }
  \end{figure}

Finally, we note that for small values of the Hubble constant $\omega$  we may use the expansions 
\begin{eqnarray}
\delta t(t)&=&\delta t'\frac{E}{m}\, -\omega\delta t'\left( \frac{\vec{P}}{E}\cdot \vec{\xi}+t'\frac{{P}^2}{E m }\right)+{\cal O}(\omega^2)\,,\\
\delta\vec{x}(t)&=&\delta\vec{x}\,'-\left(1-\frac{m}{E}\right) (\vec{n}_P\cdot \delta{\vec{x}\,}')\, \vec{n}_P\,\nonumber\\
&&+\,\omega\left(\frac{\vec{P}}{E}\,(\vec{\xi}\cdot\delta \vec{x}\,') +(t'-\vec{n}_P\cdot\vec{\xi})\frac{m P^2}{E^3}\, \delta\vec{x}\,'\right)+{\cal O}(\omega^2)
\end{eqnarray}
that hold only for small time intervals such that  $\omega \delta t\ll 1$. 

The examples we studied here show how interesting may be the kinematics of the free motion on the de Sitter spacetime. However, here we considered only simple particular examples but we believe that it deserves to investigate this whole complex phenomenology looking for new observable effects in our expanding universe.


\begin{thebibliography}{}



\bibitem{SW}
S. Weinberg, {\em Gravitation and Cosmology: Principles and Applications of the General Theory of relativity} (J. Wiley \& Sons, New York 1972). 

\bibitem{Wig}
E. Wigner, {\em Ann. Math.} {\bf 40}  (1939) 149. 

\bibitem{Mc}
G. Mackey, {\em Ann. Math.} {\bf 44}  (1942) 101.

\bibitem{WKT}
W.-K. Tung,  {\em Group Theory in Physics}  (World Sci., Philadelphia, 1984).



\bibitem{C1}
I. I. Cot\u aescu, {\em GRG} {\bf 43} (2011) 1639.

\bibitem{ES}
I. I. Cot\u aescu, {\em J. Phys. A: Math. Gen.} {\bf 33}  (2000) 9177.

\bibitem{C2}
I. I. Cot\u aescu, {\em Mod. Phys. Lett. A}  {\bf 28}  (2013) 1350033.

\bibitem{C3}
I. I. Cot\u aescu, {\em  EPL} {\bf 109}  (2015) 40002. 

\bibitem{C4}%isometry generators and canonical quantization on de Sitter spacetimes
I. I. Cot\u aescu,  arXiv:1602.06810.

\bibitem{Cnew}
I. I. Cot\u aescu,  arXiv:1701.08499.

\end{thebibliography}
\end{document}